\documentclass[12pt,a4paper]{iopart}

\usepackage{graphicx}
\usepackage{bm}% bold math
\usepackage{epsfig}
\usepackage{amssymb}
\usepackage{amsfonts}

\newcommand{\eV}{{\rm eV}}

\newcommand{\MeV}{{\rm MeV}}
\newcommand{\GeV}{{\rm GeV}}
\newcommand{\TeV}{{\rm TeV}}

\newcommand{\EeV}{{\rm EeV}}

\newcommand{\Mpc}{{\rm Mpc}}
\newcommand{\kpc}{{\rm kpc}}

\newcommand{\cm}{{\rm cm}}

\newcommand{\muG}{\mu{\rm G}}

\newcommand{\Mpl}{M_{\rm Pl}}

\graphicspath{{PLOTS/}}

\begin{document}
\hspace{11cm} {DESY 09-150}

\title{Planck-scale Lorentz violation constrained by Ultra-High-Energy Cosmic Rays}

\author{Luca Maccione}
\address{DESY, Theory Group, Notkestrasse 85, D-22607 Hamburg, Germany}
\address{{II}. Institut f\"ur Theoretische Physik, Universit\"at Hamburg, Luruper
Chaussee 149, D-22761 Hamburg, Germany}

\author{Andrew M.~Taylor}
\address{Max-Planck-Institut f\"ur Kernphysik, Saupfercheckweg, 1, D-69117, Heidelberg, Germany}

\author{David M.~Mattingly}
\address{ }

\author{Stefano Liberati}
\address{SISSA, Via Beirut, 2-4,
I-34014, Trieste, Italy}
 \address{INFN, Sezione di Trieste, Via
Valerio, 2, I-34127, Trieste, Italy}

\eads{
\mailto{luca.maccione@desy.de},
\mailto{andrew.taylor@mpi-hd.mpg.de},
\mailto{davidmmattingly@comcast.net},
\mailto{liberati@sissa.it}}

\begin{abstract}
We investigate the consequences of higher dimension Lorentz
violating, CPT even kinetic operators that couple standard model
fields to a non-zero vector field in an Effective Field Theory framework.  Comparing the ultra-high energy
cosmic ray spectrum reconstructed in the presence of such terms with data from the Pierre Auger observatory allows us to establish two sided bounds on the
coefficients of the mass dimension five and six operators for the
proton and pion.  Our bounds imply that for both protons and pions, the
energy scale of Lorentz symmetry breaking must be well above the Planck
scale. In particular, the dimension five operators are constrained at
the level of $10^{-3} M_{\rm Planck}^{-1}$.  The magnitude of the
dimension six proton coefficient is bounded at the level of $10^{-6}
M_{\rm Planck}^{-2}$ except in a narrow range where the pion and proton
coefficients are both negative and nearly equal. In this small area,
the magnitude of the dimension six proton coefficient must only be
below $10^{-3} M_{\rm Planck}^{-2}$. Constraints on the dimension six
pion coefficient are found to be much weaker, but still below
$M_{\rm Planck}^{-2}$.  
\end{abstract}

\maketitle

\section{Introduction}

Over the last decade there has been consistent theoretical interest
in possible high energy violations of local Lorentz Invariance (LI)
as well as a flourishing of observational tests. The theoretical
interest is driven primarily by hints from Quantum Gravity (QG)
ideas that local Lorentz invariance may not be an exact symmetry of
the vacuum. The possibility of outright Lorentz symmetry violation
(LV) or a different realization of the symmetry than in special
relativity has arisen in string theory~\cite{KS89, Ellis:2008gg,Horava:2009uw},
Loop QG~\cite{LoopQG,Rovelli:2002vp,Alfaro:2002ya}, non-commutative
geometry~\cite{Carroll:2001ws, Lukierski:1993wx,
AmelinoCamelia:1999pm, Chaichian:2004za}, space-time
foam~\cite{AmelinoCamelia:1997gz}, some brane-world
backgrounds~\cite{Burgess:2002tb} and condensed matter analogues of
``emergent gravity''~\cite{Analogues}.

Lorentz symmetry breaking is certainly not a necessary feature of
QG, but any Planck-scale induced LV effects could provide an observational window into QG phenomena. Moreover, the absence of LV phenomena provides by itself constraints on viable QG theories and
more firmly establishes the validity of special relativity.
Unfortunately, it is difficult to directly connect a theory of QG at
the Planck scale with low energy, testable physics. To see the difficulty from the traditional standpoint, consider
gravity as just an effective field theory
(EFT)~\cite{Burgess:2003jk} and simply quantize the spin-2 graviton
coupled to the standard model. (This approach is in contrast with
large extra dimensions which may have a much lower QG
scale~\cite{add}.) In the EFT approach there must be new quantum
gravitational effects to preserve unitarity~\cite{Han:2004wt},
however the scale of the breakdown of the theory occurs when the
center of mass energy in a scattering process nears the Planck scale
$\Mpl=1.22\times10^{19}~\GeV$. This is 15 orders of magnitude higher
than what we can directly probe at the LHC with its center of mass
beam energy of roughly 10~TeV. Therefore directly or indirectly
probing QG with a scattering or other experiment seems out of reach,
unless one is in a large extra dimensions scenario. Note however
that this scattering argument relies on the Lorentz symmetry of the
low energy effective field theory (EFT) - the meaningful Lorentz
invariant physical quantity that controls the sensitivity to physics
at $\Mpl$ is the center of mass energy. Quantities that are not LI,
such as the energy of a single particle, are irrelevant when asking
how QG affects our LI observables, in this case the scattering
amplitudes.

On the other hand, if we are specifically testing LI, the situation
changes. Here, new quantities must be introduced to describe the
physically meaningful LV physics. In particular, not only LI
quantities such as particle mass or center of mass energy are
considered in defining an observable, but also perhaps LV quantities
such as the energy of a particle in some frame, a cosmological
propagation distance, etc. These quantities can be enormous,
offsetting the tiny Planck scale in a physical observable, thereby
magnifying very small corrections (see e.g.~\cite{Mattingly:2005re,
AmelinoCamelia:2002dx,Bertolami:2000qa}). These LV quantities provide leverage and have been referred to as ``windows on QG''.

Placing these windows in a well defined framework is vital. The
standard approach is to construct a Lagrangian containing the
standard model operators and all LV operators of interest\footnote{There are other approaches to either violate or modify Lorentz invariance, that do not yield a low energy EFT (see  \cite{AmelinoCamelia:2008qg} and ref.s therein).  However, these models do not easily lend themselves to UHECR constraints as the dynamics of particles is less well understood and hence we do not consider them here.}. All renormalizable LV operators that can be added to the standard model
are known as the (minimal) Standard Model Extension
(mSME)~\cite{Colladay:1998fq}. These operators all have mass
dimension three or four and can be further classified by their
behavior under CPT. The CPT odd dimension five kinetic terms for QED
were written down in~\cite{Myers:2003fd} while the full set of
dimension five operators were analyzed in~\cite{Bolokhov:2007yc}.
The dimension five and six CPT even kinetic terms for QED for
particles coupled to a non-zero background vector, which we are
primarily interested in here, were partially analyzed
in~\cite{Mattingly:2008pw}. It is notable that SUSY forbids
renormalizable operators for matter coupled to non-zero
vectors~\cite{GrootNibbelink:2004za} but permits certain
nonrenormalizable operators at mass dimension five and six.

Many of the operators in these various EFT parameterizations of LV
have been very tightly constrained via direct observations
(see~\cite{Mattingly:2005re} for a review). The exceptions are the
dimension five and six CPT even operators, where the LV
modifications to the free particle equations of motion are
suppressed by small ratios such as $m/\Mpl$ or $E^2/\Mpl^2$, where
$m$ and $E$ are the particle mass and energy, respectively. All
operators can be tightly, albeit indirectly, constrained by EFT
arguments~\cite{Collins:2004bp} as higher dimension LV operators
induce large renormalizable ones if we assume no other relevant
physics enters between the TeV and $\Mpl$ energies. This is a very
powerful argument and should not be arbitrarily discounted. However, since it is generically expected that new physics may come into play above the TeV scale, this assumption may fail, hence the hierarchy of terms can change. Therefore, it would be nice, if possible, to constrain the dimension five and six LV CPT even kinetic terms directly via observation. This is the purpose of the present work.

How might one do this?  As mentioned, the LV corrections for these
operators are suppressed by $m/\Mpl$ or $E^2/\Mpl^2$ relative to the
LI operators. Hence one would need a very high energy particle or
very sensitive experiment to minimize this suppression. The highest
energy particles presently observed are ultra high energy cosmic
rays (UHECRs). The construction and successful operation of the
Pierre Auger Observatory (PAO) has brought UHECRs to the interest of
a wide community of scientists. Indeed, this instrument will allow,
in the near future, to assess several problems of UHECR physics 
and also to test fundamental physics with unprecedented precision
\cite{Galaverni:2007tq,Maccione:2008iw,Galaverni:2008yj}. As we shall show, it also
currently provides an extremely accurate test of Lorentz symmetry
following the introduction of these unconstrained operators.

In the past, there have been attempts to use UHECRs as a tool to
test scenarios of QG. In particular, the consequences of some
realizations of Loop QG were considered in \cite{Alfaro:2002ya}, while a
pure phenomenological and simplified approach was taken by
\cite{Aloisio:2000cm,Stecker:2004xm,GonzalezMestres:2009di}.  Recent studies analyze one of the CPT even dimension four operators (that yield a limiting
speed difference between protons and pions) in terms of the UHECR
spectrum~\cite{Scully:2008jp,Bi:2008yx}. In this work we study the
consequences of LV induced by the inclusion of CPT even dimension
five and six terms in the QED Lagrangian on the UHECR spectrum with
energies $E > 10^{19}~\eV$. By comparing the theoretical reconstructed spectrum
with the PAO observed spectrum we derive constraints on the pion and
proton dimension six LV coefficients.

This paper is structured as follows. In section~\ref{sec:theo} we outline the LV theoretical framework we adopt and the assumptions we make in this study. In section~\ref{sec:CR} we describe the present observational and theoretical status of Cosmic Ray physics. Furthermore, we describe in section~\ref{sec:LV} the effects of LV on the main processes involved in the propagation of UHECRs, while in section~\ref{sec:MC} we show the UHECR spectra resulting from our MonteCarlo simulations. Section~\ref{sec:constraints} is devoted to the presentation of the constraints we obtain on the considered LV parameters. Finally, we draw our conclusions.

\section{Theoretical framework}
\label{sec:theo}

In order to study the phenomenological consequences of LV induced by
QG, the existence of a dynamical framework in which to compute
reactions and reaction rates is essential.  We assume that the low
energy effects of LV induced by QG can be parameterized in terms of
a local EFT\footnote{In effect we assume that QG effects decouple and that at low energies they are a perturbation to the standard model + general relativity.}. Furthermore, we assume that only boost invariance is broken, while rotations are preserved (see \cite{Mattingly:2005re} for further comments on rotation breaking in this context). Therefore we introduce LV by coupling standard model
fields to a non-zero vector.

We focus on the CPT even mass dimension five and six operators
involving a vector field $u^{\alpha}$ (which we assume to describe the preferred reference frame in which the CMB is seen as isotropic), fermions (whose mass we label $m$) and
photons, that are quadratic in matter fields and hence modify the
free field equations.  The Lagrangian for a particular species of
Dirac fermion is then the usual Dirac term plus
\begin{eqnarray} \label{eq:actionfermion}
\overline {\psi}\bigg{[} - \frac {1} {\Mpl} (u \cdot D)^2
(\alpha^{(5)}_L P_L  + \alpha^{(5)}_R P_R) \\
\nonumber - \frac {i} {\Mpl^2} (u \cdot D)^3 (u \cdot \gamma)
(\alpha^{(6)}_L P_L +
\alpha^{(6)}_R P_R)  \\
\nonumber - \frac {i} {\Mpl^2} (u \cdot D) \Box (u \cdot \gamma)
(\tilde{\alpha}^{(6)}_L P_L + \tilde{\alpha}^{(6)}_R P_R) \bigg{]}
\psi
\end{eqnarray}
where $u^a$ is a timelike unit vector describing the preferred
frame, $P_R$ and $P_L$ are the usual right and left projection
operators, $P_{R,L}=(1 \pm \gamma^5)/2$, and $D$ is the gauge
covariant derivative.  The $\alpha$ coefficients are dimensionless.
The additional photon operator is
\begin{equation} \label{eq:actionphoton}
 - \frac {1} {2 \Mpl^2} \beta^{(6)}_\gamma F^{\mu \nu} u_\mu
u^\sigma (u \cdot \partial)^2 F_{\sigma \nu}\;.
\end{equation}
For fermions, at $E\gg m$ the helicity eigenstates are almost
chiral, with mixing due to the particle mass and the dimension five
operators. Since we will be interested in high energy states, we
re-label the $\alpha$ coefficients by helicity, i.e.
$\alpha^{(d)}_{+}=\alpha^{(d)}_{R},
\alpha^{(d)}_{-}=\alpha^{(d)}_{L}$. The resulting high energy
dispersion relation for positive and negative helicity particles can
easily be seen from (\ref{eq:actionfermion}) to involve only the
appropriate $\alpha^{(d)}_{+}$ or $\alpha^{(d)}_{-}$ terms.  For
compactness, we denote the helicity based dispersion by
$\alpha^{(d)}_{\pm}$.  Therefore at high energies we have the
dispersion relation (see also \cite{Glinka:2008tr})
\begin{equation} \label{eq:dispfermionhighE}
E^2 =p^2+m^2 + f^{(4)}_{\pm} p^2 +f^{(6)}_{\pm} \frac{p^4} {\Mpl^2}
\end{equation}
where $f^{(4)}_{\pm}=\frac {m} {\Mpl} (\alpha^{(5)}_-  +
\alpha^{(5)}_+) $ and $f^{(6)}_{\pm}= 2\alpha^{(6)}_{\pm} +
\alpha^{(5)}_- \alpha^{(5)}_+$.  We have dropped the
$\tilde{\alpha}^{(6)}_{R,L}$ terms as the $\Box$ operator present in
these terms makes the correction to the equations of motion
proportional to $m^2$ and hence tiny.

In Lorentz gauge the photon dispersion relation is
\begin{equation} \label{eq:dispphoton}
\omega^2=k^2  +  \beta^{(6)} \frac {k^4} {\Mpl^2}\;.
\end{equation}

Before we continue, we make a simplifying assumption - that parity
is a symmetry in our framework.  In particular, this implies that
our helicity coefficients are equal. There is no underlying
motivation from QG as to why parity should be approximately valid if
LI is broken, however it is reasonable to assume this for the first
attempt at constraints. The parity violating case, which involves
helicity decay reactions in addition to the ones considered here, we
leave for future work.

The dimension five fermion operators induce two corrections, one
proportional to $E^4$ and one corresponding to a change in the
limiting speed of the fermion away from $c$. Constraints on a
different limiting speed for pions and protons in the context of
UHECR have been derived in~\cite{Scully:2008jp}, $\delta_{\pi
p}=f^{(4)}_\pi-f^{(4)}_p<10^{-23}$ if $\delta_{\pi p}>0$.
Complementing this constraint, if $\delta_{\pi p}<0$ then the
necessary absence of a vacuum \v{C}erenkov (VC) effect for high energy
protons~\cite{Jacobson:2002hd} (see section~\ref{subsec:VC} for a more detailed
discussion of the VC effect) limits $\delta_{\pi
p}>-10^{-22}$. In our parameterization with the parity assumption,
the $\alpha^{(5)}$ coefficients are therefore immediately
constrained at the $10^{-3}$ level. Hence we will drop them for the
rest of this paper and concentrate on the dimension six terms.

Since parity is conserved, $f^{(6)} \equiv f^{(6)}_+ = f^{(6)}_-$.
We define $\eta_p=f^{(6)}_p$ and $\eta_\pi=f^{(6)}_\pi$, and drop the superscript from $\beta^{(6)}$. Hence, the dispersion relations we assume in this work for protons, pions, and photons respectively, are
\begin{eqnarray}\label{eq:finaldisp}
E_p^2=p^2  + m_p^2 +  \eta_p \frac {p^4} {\Mpl^2}\nonumber\\
E_\pi^2=p^2  + m_\pi^2+ \eta_\pi \frac {p^4} {\Mpl^2}\\
\omega^2=k^2  +  \beta \frac {k^4} {\Mpl^2}\nonumber\;.
\end{eqnarray}

Although there are indications that these operators may be strongly
constrained \cite{Galaverni:2007tq,Maccione:2008iw,
Mattingly:2008pw}, nothing conclusive has been claimed yet, as high
energy particles are needed to probe the effects of these operators.
A fairly accurate general estimate of the energy range in which LV
corrections in equations (\ref{eq:finaldisp}) are relevant is
obtained by comparing the largest mass of the particles entering in
the LV reaction with the magnitude of the LV correction in these
equations~\cite{Jacobson:2002hd}. In our case, assuming
$\eta_p,\eta_\pi \sim 1$, the typical energy at which LV
contributions start to be relevant is of order
 $E_{th} \sim \sqrt{m_{p}\Mpl} \simeq 3 \times 10^{18}~\eV$, a fairly
reachable energy for UHECR experiments. Note that if one considers
neutrinos, then the typical energy (assuming, as a worst case scenario, $m_{\nu}
\simeq 1~\eV$) is $E_{th}\sim 100~\TeV$, well within reach of
neutrino telescopes such as ICECUBE or Km3NeT. However, we will
neglect them here, as even the confirmed detection of high energy
astrophysical neutrinos has not yet been
achieved~\cite{Collaboration:2008ih}. Hence, at present, only
observations in the field of UHECR physics, with energy of order $E \gtrsim 10^{19}~\eV$, can provide significant information on such type of LV.

UHECR's are, in general, assumed to be composite objects, being
either protons or nuclei. In our EFT approach, the fermionic
operators apply to the quark constituents, there are other LV
operators for gluons, and the proton LV is a combination of all the
LV for the constituents.  This is the approach taken in
\cite{Gagnon:2004xh}, where a parton model is assumed for protons
and the net proton LV is determined by the LV terms for the partons
along with the parton fraction at UHECR energies. If we really want
to establish constraints on the bare parameters in the action, we
would need to do the same type of analysis. Our goal is not so
ambitious - we will treat protons, photons, and pions as individual
particles with their own independent dispersion relations and
constrain the $\eta_p, \eta_\pi, \beta$ coefficients. This approach
is phenomenologically valid since we are using energy-momentum
conservation and the initial and final state particles are separated
composite fermions with well-defined energy, momenta, and dispersion
relations.  These composite dispersion relations are therefore what
must appear in the energy-momentum conserving $\delta$-functions in
the scattering amplitude.

It is possible, of course, that one could have LV for quarks and
none for hadrons if LV was possible only for particles with color
charge.  In this case our results would be very misleading. However, since we assume that LV comes from QG and not a modification of QCD we do not give this
possibility much credence and so will ignore it. Hence, underlying
our treatment is the assumption that CPT even dimension six
operators for the fundamental partons generate net CPT even
dimension six operators for the composite particles of the same
order. Results derived by treating every particle as independent in
this way are weaker than what one might get using a parton approach,
where many different particles are made of only a few constituents.

Now that we have our theoretical background, we turn to the UHECR
spectrum.

\section{Cosmic Ray spectrum}
\label{sec:CR}

The Cosmic Ray spectrum spans more than ten decades in energy (from
$<100~\MeV$ to $>10^{20}~\eV$) with a power-law shape of impressive
regularity
\begin{equation}
\frac{dN}{dE} \propto E^{-p}\;.
\end{equation}
The spectral slope $p$ has been measured as $p \simeq 2.7$ for $1~\GeV \lesssim E \lesssim 10^{15.5}~\eV$, followed by a softening (the ``knee'') to $p \simeq 3.0$ for $10^{15.5}~\eV \lesssim E \lesssim 10^{17.5}~\eV$, a further steepening to $p\simeq 3.2$ (the ``second knee'') up to $E \simeq 10^{18.5}~\eV$ and a subsequent hardening (the ``ankle'') to again $p \simeq 2.7$ at $E \gtrsim 10^{18.5}~\eV$ \cite{Gaisser:2006sf,Bergman:2007kn}.

One of the most puzzling problems in CR physics concerns their origin. Being charged particles, their paths are deflected in both Galactic and extragalactic magnetic fields during propagation, erasing the information about their source direction, resulting in their observed arrival directions being almost isotropically distributed. Only CRs with sufficient energy, $E \gtrsim 10^{20}~\eV$, are capable of remaining predominantly undeflected by nG extragalactic magnetic fields \cite{Hooper:2006tn}, leading to the expectation of some anisotropy, as was found experimentally \cite{Cronin:2007zz}. In fact, the expected angle $\delta$ of deflection due to, e.g., galactic magnetic fields (GMFs), is (we assume $1~\kpc$ as the typical coherence length of the GMFs and a mean field strength of $3~\muG$) \cite{Cronin:2007zz}
\begin{equation}
\delta =
2.7^{\circ}Z\left(\frac{60~\EeV}{E}\right)\left(\frac{x}{1~\kpc}\right)\left(\frac{B}{3~\muG}\right)
\end{equation}
decreasing as the proton energy $E$ increases due to the reduced
fraction of the proton's energy in the field.

A second CR puzzle regards the ``cross-over'' energy at which the
sources of the cosmic rays detected at Earth change from being
predominantly Galactic to extragalactic. Interestingly, lower limit
constraints have been placed on this transition energy using the
ultra-high energy neutrino flux observation limits set by AMANDA
observations \cite{Ahlers:2005sn,Ahlers:2009rf}, with too low a transition energy
requiring too large an energy budget for extragalactic sources,
resulting in the possibility of the expected ultra-high energy
neutrino fluxes being in conflict with the observational limits
(under certain assumptions about the source acceleration
efficiency, evolution with redshift, and UHECR spectral index). However, the energy of the transition to a dominance of
extragalactic cosmic ray sources continues to remain unclear. It is
natural to expect a flattening of the energy spectrum at the
transition energy, with the harder subdominant extragalactic
component taking over from the softer Galactic component. In this
respect, associating the ``ankle'' feature with the cross-over
energy certainly provides a coherent picture for the transition. At
this energy the (proton) Larmor radius in the Galaxy's $\mu$G field
begins to exceed the thickness of the Milky Way disk and one expects
the Galactic component of the spectrum to die out. Subsequently, the
end-point of the Galactic flux ought to be dominated by heavy
nuclei, as these have a smaller Larmor radius for a given energy,
and some data is indeed consistent with a transition from heavy
nuclei to a lighter composition at the ankle \cite{Allard:2005cx,Berezinsky:2007wf,Unger:2007mc}.

A third puzzle regarding CRs is at what energy the end-point to the
CR spectrum occurs. A suppression to the spectrum is naturally
expected theoretically due to the interactions of UHECR protons with
the Cosmic Microwave Background (CMB).  This interaction leads to
the production of charged and neutral pions, eventually dumping the
energy of the UHECR protons into neutrinos and $\gamma$-rays. At the present epoch, significant photo-pion production in a LI theory occurs only if the energy of the interacting proton is above $10^{19.6}~\eV$, with a rapid decrease in their mean-free-path above
this energy. Hence, it has long been thought to be responsible for a
cut-off in the UHECR spectrum, the Greisen-Zatsepin-Kuzmin (GZK)
cut-off \cite{gzk}. Moreover, trans-GZK particles arriving at Earth
must be accelerated within the so called GZK sphere, whose radius is
expected to be of the order of 100~Mpc at $\sim 10^{20}~\eV$ and to
shrink down at larger energies. A simple analytic description of the
GZK sphere can be given (see~\ref{analytic}),
\begin{equation}
l_{\rm horiz.}=\frac{l_{0}}{\left[e^{-x}(1-e^{-x})\right]}\;,
\end{equation}
with $l_{0} = 5~\Mpc$ and $x\sim 3.4\times10^{20}~\eV/E_{p}$.
Experimentally, the presence of a suppression of the UHECR spectrum
has been confirmed only recently with the observations by the HiRes
detector \cite{Abbasi:2007sv} and the PAO
\cite{Roth:2007in}. Although the cut-off could be also due to the
finite acceleration power of the UHECR sources, the fact that it
occurs at roughly the expected energy favors a GZK explanation. The
correlation results shown in \cite{Cronin:2007zz} further strengthen this
hypothesis. It is this last puzzle where possible LV effects come
into play.

%%%%%%%%%%%%%%%%%%%%%%%%%%%%%%%%%%%%%%%%
%%%%%%%%%%%%%%%%%%%%%%%%%%%%%%%%%%%%%%%%

\section{UHECR Proton Interactions and LV}
\label{sec:LV}

\begin{figure}[tbp]
   \centering\leavevmode
   \includegraphics[width=0.4\linewidth,angle=-90]{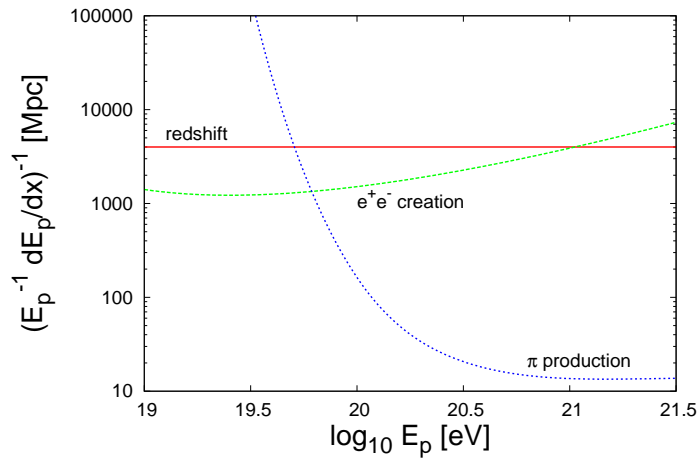} % requires the graphicx package
   \caption{Comparison of the typical time scales for the processes relevant for UHECR proton propagation. Only interaction with CMB photons have been considered here.}
   \label{fig:timescale-comp}
\end{figure}

As they propagate from their source to Earth, UHECRs lose their energy in several ways. Besides adiabatic losses due to the expansion of the Universe, whose LV modifications will be neglected in the following, the most relevant energy loss mechanisms for protons are pair production through interactions with the CMB (dominant in the present epoch for $E_{p}< 10^{19.6}~\eV$) and photo-pion production through interactions with the CMB (dominant in the present epoch for $E_{p}>10^{19.6}~\eV$). The typical loss time-scales for these processes are shown in Fig.~\ref{fig:timescale-comp}. 

The effect of LV on UHECR propagation is twofold: it modifies
standard reactions and allows new, normally forbidden reactions. In particular, in the following subsection we will consider 
\begin{itemize}
\item $p+\gamma \rightarrow p+\pi^{0} ~(n+\pi^{+})$, which is modified by LV.
\item $p\rightarrow p+\gamma$ and  $p\rightarrow p+\pi$, which correspond respectively to photon and pion emission in vacuum and would be forbidden if LI were exact.
\end{itemize}

Before moving to a detailed description of these processes in a LV framework, it is worth discussing the role played by the possible presence of nuclei in UHECRs, as supported by increasing experimental evidence \cite{Unger:2007mc}.
Let us consider here how our proton LV terms also affect UHECR nuclei propagation. At Earth, a given cosmic ray heavy nuclei has a certain total momentum $p_N$.  Assuming the total momentum is equally distributed amongst the constituent nucleons, each nucleon possesses a momentum $p_N/A$, where $A$ is the mass number of the nucleus.  A iron nuclei ($A=56$) at $10^{20}$ eV then has nucleons with momenta only at $2 \times 10^{18}$ eV.  This is much lower than the individual protons we are considering at momenta $>10^{19}$ eV, and since our LV scales heavily with momenta the propagation of heavy nuclei is largely
unaffected by LV. We assume therefore the interactions of heavy nuclei may be treated as if LI was still exact.

Since nuclei propagation remains unaffected by the LV terms
discussed in this paper, the results from previous work on nuclei
propagation such as \cite{Hooper:2006tn,Hooper:2008pm,Allard:2008gj} remain
applicable here. The general conclusion from this work is that a
variety of source compositions, from protons to iron nuclei, can
be consistently assumed to be the sole injection composition at each CR
source, without being in conflict with either the CR spectrum or
elongation rate data.

However, in order to get clear constraints and in agreement with the evidence on the anisotropic distribution for UHECR recently reported by AUGER \cite{Cronin:2007zz}, we shall assume here a purely protonic flux at the energies of our interests.

\subsection{Modified GZK}
In a LI theory, photo-pion production $p+\gamma \rightarrow
p+\pi^{0}(n+\pi^{+})$ is the highest energy-loss process that occurs
during the propagation of UHECR protons. It is therefore crucial to
carefully investigate how LV affects its characteristics. The most
important quantities needed to compute the UHECR spectrum are the
mean-free-path $\lambda$ of protons and the fraction of initial
proton momentum transferred to the outgoing pion, the so called
inelasticity $y$.

\subsubsection{LV mean-free-path}
We want to calculate the mean-free-path $\lambda$ for a proton
undergoing GZK interactions with the CMB. We assume here that LV
does not strongly affect the dynamics of the photo-pion production,
hence that the LV cross section is roughly equal to the LI one,
apart from small corrections that we neglect. We discuss below the
potential effects of LV on the cross-section. Assuming the LV
dispersion relations outlined in Eq.~(\ref{eq:finaldisp}), with
$\beta = 0$, the mean-free-path $\lambda$ can be calculated as
\begin{equation}
\lambda^{-1} = \int_{\epsilon_{\rm
min}%(E_{p},\eta_{p},\eta_{\pi})
}^{\infty}{\rm d}\epsilon\int_{-1}^{1}\frac{{\rm d}\cos\theta}{2}
n(\epsilon)\sigma(s)\left(1-v_{p}\cos\theta\right) \label{eq:lambda}
\end{equation}
where $\epsilon$ is the energy of the incoming photon, $n(\epsilon)$
is the number density of the target photons (which are isotropically
distributed in space), the photon threshold energy $\epsilon_{\rm
min}$ depends in general on $E_{p}$, $\eta_{p}$ and $\eta_{\pi}$,
$\sigma(s)$ is the total cross-section, dependent on the ``square
center of mass energy'' $s = (p_{p}+p_{\gamma})^{2}$, $v_{p}$ is the
velocity of the proton with energy $E_{p}$ and $\theta$ is the angle
between the direction of the incoming photon and that of the
incoming proton.

According to our definition, we can write
\begin{equation}
s \equiv (p_{p}+p_{\gamma})^{2} \simeq m_{p}^{2} +
2p\epsilon(1-\cos(\theta)) + \eta_{p}\frac{p^{4}}{\Mpl^{2}}
\end{equation}
neglecting terms of order  $\epsilon/p \ll 1$.
For UHECR, even with LV, $v_p$ is extremely close to one for any reasonable value of $\eta_p$ and therefore we set it exactly equal to one. 
Hence, we can re-express (\ref{eq:lambda}) as
\begin{equation}
\lambda^{-1} = \frac{1}{8p^{2}} \int_{\epsilon_{\rm
min}%(E_{p},\eta_{p},\eta_{\pi})
}^{\infty} {\rm d}\epsilon\,\frac{n(\epsilon)}{\epsilon^{2}}
\int_{s_{\rm min}}^{s_{\rm max}} {\rm d}s
\left(s-s_{\rm min}
\right)\sigma(s) \label{mean-free-path}
\end{equation}
where $s_{\rm min} = m_{p}^{2}+\eta_{p}E_{p}^{4}/\Mpl^{2}$
and $s_{\rm max}(\epsilon) = s_{\rm min} + 4E_{p}\epsilon$
\cite{Mucke:1999yb}.

The threshold values $\epsilon_{\rm min}(E_{p},\eta_{p},\eta_{\pi})$
correspond to the solution of the energy-momentum conservation
equation in the threshold configuration \cite{Mattingly:2002ba}
\begin{equation}
\fl
4p\epsilon y(1-y)-m_p^2y^2-m_\pi^2(1-y) +
\frac{p^4}{\Mpl^2}y(1-y)\left[\eta_p(1-(1-y)^3)-\eta_\pi y^3 \right]
= 0\;, \label{eq:threshold}
\end{equation}
where $y = p_{\pi}/p$ is the inelasticity. In order to compute the
LV threshold energy, we solve numerically Eq.~(\ref{eq:threshold}).

\subsubsection{Comments on phase space effects}

In the computation above we neglected direct contributions to the
total cross-section coming from LV. The total cross-section is
calculated as
\begin{equation}
\sigma(s) = \int_{x_{\rm min}}^{x_{\rm max}} dx\frac{d\sigma}{dx}
\end{equation}
where $x = \cos\theta$ and $\theta$ is the angle between the
incoming and the outgoing proton. This quantity is related to the LI
Mandelstam variable $t = (p_{\rm p,in}-p_{\rm p,out})^{2} = (p_{\pi}
- p_{\gamma})^{2}$.

In order to evaluate LV corrections to the total cross-section we
have to consider different possible contributions from both
kinematics and dynamics. While we do not expect dynamical
contributions (i.e.~from $|{\mathcal M}|^{2}$) to be relevant,
because, by analogy with findings in LV QED \cite{Jacobson:2005bg},
they are Planck-suppressed with respect to ordinary ones,
corrections to the kinematics could in principle play an important
r\^ole. However, in the LI case the differential cross-section is
known to be strongly peaked at $\cos\theta \simeq 1$, i.e.~in the
forward direction, with an exponential suppression of
high-transverse momentum production \cite{Mucke:1999yb}, which is
usually modeled, for small values of $|t|$, as
\begin{equation}
\frac{d\sigma}{dt} = \sigma_{0}e^{bt}\;,
\end{equation}
where $b \simeq 12~\GeV^{-2}$ as determined experimentally (notice
that $t<0$ by construction\footnote{The careful reader might be worried by the fact that this is no longer ensured in LV physics, hence one could have $t\geq0$ at some energy for some combination of the LV parameters. However, we notice that the condition $t=0$ sets the onset of the process of \v{C}erenkov emission in vacuum (which we discuss in section~\ref{subsec:VC}). Since for each combination of LV parameters we consider the GZK reaction only at energies below the VC threshold, the condition $t<0$ is guaranteed.}). 

We expect then that only LV corrections affecting the behavior of
$\cos\theta$ near $\cos\theta\simeq1$ are important for our estimate
of the total cross section $\sigma$, being other corrections
exponentially suppressed. In order to estimate how $\cos\theta$ is
affected by LV physics, we numerically compare the expectation
values of $\cos\theta$ in both LI and LV cases for various
configurations of the interacting particles.
We find that LV contributions are indeed relevant in the region
$\cos\theta \simeq 1$. However, we notice that neglecting LV effects
in the cross-section is a conservative approach. In fact, it is
possible to show that the way LV affects the cross-section is such
that it enhances distortions from the LI GZK process. Indeed, when
the threshold energy is lowered (hence, protons are able to interact
with more photons) the cross-section is increased (hence, the
probability of interaction is enhanced as well), while when the
threshold energy is increased the cross-section is lowered.
Therefore, neglecting LV effects in the cross-section amounts to
underestimating LV effects in UHECR proton propagation, thereby
implying conservative limits.

\subsubsection{LV inelasticity}

The other important quantity entering in the computation of the
UHECR spectrum is the proton {\em attenuation length} for photo-pion
production onto the radiation background. The attenuation length
expresses the mean distance over which a proton must travel to
reduce its energy to $1/e$ of its initial one and is usually defined
as \cite{Stecker:1968uc}
\begin{equation}
\frac{1}{E_{p}}\frac{dE_{p}}{dx} = \frac{1}{8p_{p}^{2}}
\int_{\epsilon_{\rm min}(E_{p},\eta_{p},\eta_{\pi})}^{\infty} {\rm
d}\epsilon\,\frac{n(\epsilon)}{\epsilon^{2}}  \int_{s_{\rm
min}}^{s_{\rm max}(\epsilon)} {\rm d}s
\left(s-s_{\rm min}
\right)y(s)\,\sigma(s)\;,
\end{equation}
where
\begin{equation}
y(s) = \frac{1}{2}\left(1-\frac{m_{p}^{2}-m_{\pi}^{2}}{s}\right)
\label{eq:inelasticity}
\end{equation}
is the inelasticity of the process as computed in the LI case for
the single pion emission process. At threshold, $s_{\rm
th}=(m_{p}+m_{\pi})^{2}$, giving, $y(s_{\rm th})\approx~0.13$. At
energies well above threshold, the multiplicity of the photo-pion
production process grows and the above equation for the inelasticity
no longer holds.

The computation of $y(s)$ is an issue, since we need to compute the energy-momentum conservation also in off-threshold configurations. However, since LV corrections to $y(s)$ are relevant only near
threshold (in the LV case $y(s) \rightarrow 1/2$ as $s$ increases as
well) we assume that Eq.~(\ref{eq:inelasticity}) is valid for $s
\gtrsim s_{\rm th}$.

The problem is then reduced to what to assume for $y(s)$ around
$s_{\rm th}$, where the LV corrections are in principle important.
As for the total cross-section, we assume that the analytic
expression (\ref{eq:inelasticity}) is not modified provided $s$ is
computed taking into account LV. In order to check our assumption,
we notice that we are able to compute easily the expected value
$\bar{y}$ of $y$ at threshold, because the solution for $p$ of the
energy-momentum conservation (\ref{eq:threshold}), together with the
requirement that $p$ is minimum, provides us with the pair $(p_{\rm
th}, \bar{y})$, or equivalently $(s_{\rm th}, \bar{y})$. We find that, as long as $s_{th} \gtrsim m_{p}^{2}$, this procedure is valid, as the values $y(s_{\rm th})$ obtained by extrapolating Eq.~(\ref{eq:inelasticity}) down to $s_{\rm th}$ are well in agreement, within $10^{-3}$, with $\bar{y}$. If instead $s_{\rm th} \lesssim m_{p}^{2}$, then the numerically evaluated inelasticity may differ significantly from the extrapolated one. However, in this case the inelasticity is dramatically reduced (or dramatically increased, to $y \sim 1$, when $s < 0$), compared to the LI one, reaching values of $y < 0.005$. When this happens, protons do not lose their energy effectively (or they do lose most of it in just one interaction) during propagation, which leads to clear inconsistencies with experimental observations, as will be shown below.

\subsection{Vacuum \v{C}erenkov emission}
\label{subsec:VC}

LV allows two more processes competing with the photo-pion-production to be active: photon and pion VC emission. In fact, due to LV a proton can spontaneously emit photons (or neutral pions) without violating energy-momentum conservation.

It has been shown in other contexts (LV QED
\cite{Jacobson:2002hd,Jacobson:2005bg}) that the reaction rate for
such processes is of the order of a nanosecond, acting then as a
sharp effective cut-off on the particle spectrum.   We follow
the same analysis as in~\cite{Jacobson:2005bg}, for pion as well as photon emission, but considering our operators. For the case of pion emission we use the Yukawa nucleon-pion
matrix element. A straightforward calculation shows that the VC
rates for both photons and pions become extremely fast very quickly
above threshold. Hence computing the threshold energies of both
processes and cutting off the spectrum at those energies is
sufficient for our aims, the typical VC time scales being many
orders of magnitude shorter than the time scales of the other
processes involved in UHECR propagation. We implement the cut-off by
setting the attenuation length for particles above VC threshold to
the value $c\times 1~{\rm ns} \simeq 30~\cm$.

Let us first discuss VC with emitted photons, as it will be the
simpler case.  The threshold energy depends, in general, on both
$\eta_p$ and $\beta$.   Our goal, however is to constrain $\eta_p$
and $\eta_\pi$, not $\beta$.  Hence we need a simplification such
that $\beta$ becomes irrelevant, which will allow us to place a
constraint only on $\eta_p$. We can achieve such a simplification by
considering only low energy photon emission. Since the dispersion
correction scales as $k^4$, LV is irrelevant for low energy photons
unless $\beta$ is unnaturally large.  Hence we can ignore $\beta$
for soft photon emission.  The photon VC effect then becomes very
similar to the ordinary \v{C}erenkov effect, where there is
\v{C}erenkov emission when the group velocity of a particle exceeds
the low energy speed of light.  This happens at some UHECR momenta
provided $\eta_p>0$.

One might be concerned that considering only low energy photon
emission, which is a small part of the outgoing phase space, would
give a rate that is too low to give our sharp cut-off. This can be
shown explicitly to not be the case. For example, if we assume
$\eta_p$ of O(1) and $\beta<10^8$, which is unnaturally large, then
we can neglect $\beta$ for photon energies up to one-hundredth the
initial proton energy.  This provides easily enough phase space to
yield a high rate directly above threshold, justifying the cut-off
implementation mentioned above.  We therefore impose a cut-off to
the UHECR spectrum in the $\eta_p>0$ half-plane at momentum
\begin{equation}
p_{\gamma VC} =
\left(\frac{m_{p}^{2}\Mpl^{2}}{3\eta_{p}}\right)^{1/4} \qquad\qquad
\eta_{p}>0\;. 
\label{eq:pgVC}
\end{equation}

On the other hand, we treat pion VC differently.  We want to limit
both $\eta_p$ and $\eta_\pi$, hence we will consider both hard and
soft pion emission. This means that we use the whole outgoing phase
space. However, we lose the ability to analytically solve the threshold
equations. Instead, the pion VC threshold energy has to be computed
numerically as
\begin{equation}
\frac{p_{\pi VC}}{(\Mpl m_{p})^{1/2}} = \min_{y\in (0,1)}
\left(\frac{ 1/(1-y) + m_{\pi}^{2}/(m_{p}^{2} y^{2})}{
\eta_{p}(y^{2}-3y+3) - \eta_{\pi}y^{2}(2y+1)}\right)^{1/4}\;,
\label{eq:ppVC}
\end{equation}
where, as usual, $y$ is the fraction of initial momentum going to
the pion.  Note that where this equation has no real solution, pion
VC does not occur.

\section{Results}
\label{sec:MC}

\subsection{Monte Carlo simulation to obtain GZK feature}

During UHE proton propagation, the dominant energy-loss process
leading to attenuation at the highest energies occurs via photo-pion
production, $p\gamma\rightarrow p+\pi^{0}/n+\pi^{+}$ (note the pion
multiplicity is 1 for interactions close to threshold), as shown in
Fig.~\ref{fig:timescale-comp}. This process has a typical
inelasticity of the order of 20\% (see Eq.~\ref{eq:inelasticity}),
meaning that each time it interacts, a proton loses roughly $1/5$ of
its total energy. Moreover, the attenuation length of a UHECR proton
is roughly a few Mpc, as highlighted by Eq.~(\ref{eq:horizon-rough})
and seen in Fig.~\ref{fig:horizon}. UHECRs then undergo between 1
and 10 photo-pion production interactions during their journey from
source to Earth, but not substantially more. Therefore, it is not
justified to think of this energy-loss process as if it was
happening continuously. Rather, a MonteCarlo approach should be
adopted, to take into account the stochastic nature of the GZK
process.

In order to understand the main effects of LV on the UHECR spectrum
we present the results of pure proton composition of UHECRs under
the assumption of a continuous distribution of sources, distributed
as
\begin{eqnarray}
\frac{dN}{dV}\; &=& 0 \qquad\qquad\quad  0<z<z_{\rm min}\label{source-space}\\
&\propto& (1+z)^{3} \qquad z_{\rm min}<z<1.0\nonumber
\end{eqnarray}
where $dN/dV$ describes the number of sources in a comoving volume
element and $z$ is the redshift at which the source density is being
considered. The free parameter $z_{\rm min}$ is varied to
investigate the effects of the closest source, which might be non-trivial. Indeed, if LI were exact, UHECR protons in the cut-off region would only travel distances $<100~\Mpc$, hence only local ($z\ll1$) sources can actually contribute to the arriving flux at these energies. Therefore, LI spectrum reconstruction is mildly affected by the actual value of $z_{\rm min}$ as we shall see later. However, if LI is violated this conclusion could be changed, as protons may travel substantially longer distances without losing energy. Along with this, the spectrum of CRs injected by each source was assumed to be of the form,
\begin{eqnarray}
\frac{dN_{p}}{dE_{p}}\;&\propto& E^{-\alpha} \qquad E<E_{c}\label{source-energy}\\
&\propto& 0  \qquad\quad\; E>E_{c}\nonumber
\end{eqnarray}

Throughout this paper, $E_{c}=10^{21}~\eV$ and $\alpha = 2$ will be used unless stated otherwise. In what follows we investigate both
the effect on the arriving CR flux introduced by our LV terms as
well as the effect of introducing a mimimum distance to the first
source. Such a minimum distance consideration is introduced to
enable the reader to differentiate the effect this has on the GZK
feature from the effects introduced by the LV terms.

\subsection{LV effects on the cut-off feature}

By employing a Monte Carlo description for the propagation of UHECR
protons, including the effects introduced through the consideration
of the LV terms discussed, we obtain the expected fluxes arriving at
Earth following the injection of protons with spectra of the form
shown in Eq.~(\ref{source-energy}) at their sources, whose spatial
distribution is given in Eq.~(\ref{source-space}), with $z_{\rm min} = 0$.

\begin{figure}[tbp]
   \centering
   \includegraphics[scale = 0.5]{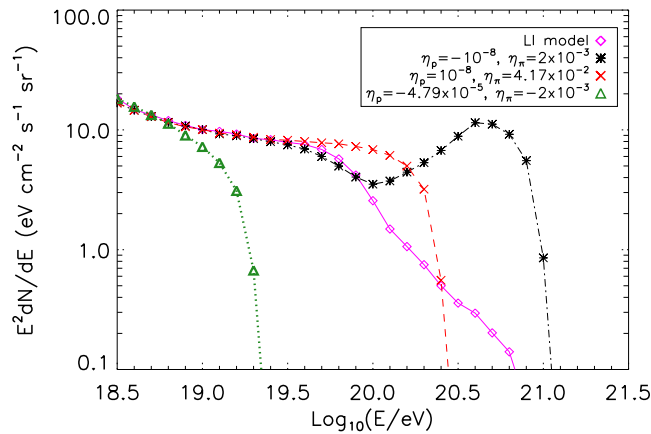}
   \caption{A range of UHECR proton spectra for different values of $(\eta_{p},\eta_{\pi})$. An injection spectrum of $\alpha=2.0$ and $E_{c}=10^{21}$~eV have been used in these calculations.}
   \label{fig:spectra-plots-range}
\end{figure}

In Fig.~\ref{fig:spectra-plots-range} we show the almost complete range of results obtainable from Monte Carlo simulations for the propagation of UHECR protons including our LV effects, for different LV parameters $\eta_{p}$ and $\eta_{\pi}$. 
The LV term effects vary from a simple complete GZK-like cut-off (with or without recovery as in the case of a GZK-suppression) of the proton flux, to a early (or delayed) onset of the cut-off to higher energies followed by a stronger cut-off when it occurs.

We now describe the main features in more detail. Firstly, the effect of VC emission is clearly evident, as seen in the green dotted curve. The VC emission acts as a sharp cut off on the UHECR spectrum. Note that since this cut-off is at the source there is not only the effect of the hard cutoff in the spectrum at $E=E_{\rm th}^{\rm VC}$, but also a suppression of the UHECR flux at lower energies due to the absence of the higher energy source protons that would have wound up with $E<E_{\rm th}^{\rm VC}$ eV at Earth due to GZK losses. 

Secondly, for the case of $\eta_{p}=10^{-8}$ and $\eta_{\pi}\sim 4\times10^{-2}$, corresponding to the red dashed curve, the GZK cut-off feature is seen to be delayed compared to the LI case, turning on very quickly at around $10^{20.3}~\eV$. Interestingly, this is a general effect seen in all cases for ``large'' $\eta_{\pi}>0$ (compared to $\eta_{p}$).  In all such cases, the cut-off feature exhibited is both initially ($\sim10^{19.6}~\eV$) delayed and very hard after turning on. Note that to understand this second effect of delay plus strong cut off considering only the effects of VC emission is not sufficient, since VC depends only on $\eta_{p}$ in the first quadrant, whereas we see that changes in $\eta_{\pi}$ affect the UHECR spectrum. The delay is easily understood, though, as positive $\eta_\pi$ increases the effective mass of the pion, thereby delaying the GZK cutoff. The cutoff is sharper since, for any given background photon energy, once the reaction occurs the phase space opens up more rapidly than in the LI case due to the scaling of the LV dispersion corrections with energy.

The black solid curve shows another important effect. In this case,
$\eta_{p}<0$, while $\eta_{\pi} > 0$. While for the chosen
combination of parameters the GZK feature turns on at nearly the LI
energy (compare the black and the magenta curve), the spectrum
exhibits a strong enhancement of the flux above $10^{20}~\eV$. The
reason is that if $\eta_{\pi}> 0$ then the effective pion mass is
increased at high energy, hence the GZK process can be effectively
inhibited. We have thus that high energy ($>10^{20}~\eV$ in the case
of Fig.~\ref{fig:spectra-plots-range}) protons are no longer
absorbed by the photon radiation fields. A similar feature of flux
recovery has been found also in \cite{Scully:2008jp}.

\subsection{Effects of distance to the closest source on the cut-off feature}

\begin{figure}[tbp]
   \centering\leavevmode
   \includegraphics[scale = 0.5]{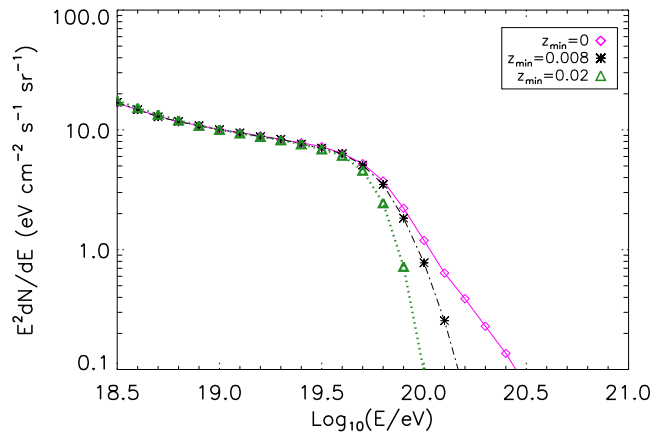}
   \caption{Comparison of spectra of UHECRs (pure protons) obtained with different values of $z_{\rm min}$. An injection spectrum of $\alpha$=2.0 and $E_{c}$=10$^{21}$~eV have been used in these calculations.}
   \label{fig:spectra-plots-zmin}
\end{figure}

In order to distinguish the effects that LV terms may introduce to
the GZK cut-off feature, we here consider the effects on this
cut-off feature introduced by non-zero values of $z_{\rm min}$ on a LI spectrum model. In Fig.~\ref{fig:spectra-plots-zmin} we show the results for the
spectra obtained using different $z_{\rm min}$. By increasing the distance
between the first UHECR source and Earth, the high energy GZK
feature is seen to become much steeper as has been demonstrated
previously in \cite{Berezinsky:2002nc}. With the effects introduced
by the existence of a non-zero $z_{\rm min}$ in mind, the
differences this introduces into the shape of the cut-off feature
compared to that introduced by LV terms are demonstrated to be quite
distinct, with LV terms typically leading to a harder cut-off in the
energy spectrum than usually expected from LI calculations.

\section{Constraints from UHECR observations}
\label{sec:constraints}

UHECR observations indeed provide strong constraints on the available LV parameter space.

\begin{figure}[tbp]
   \centering\leavevmode
   \includegraphics[scale = 0.5]{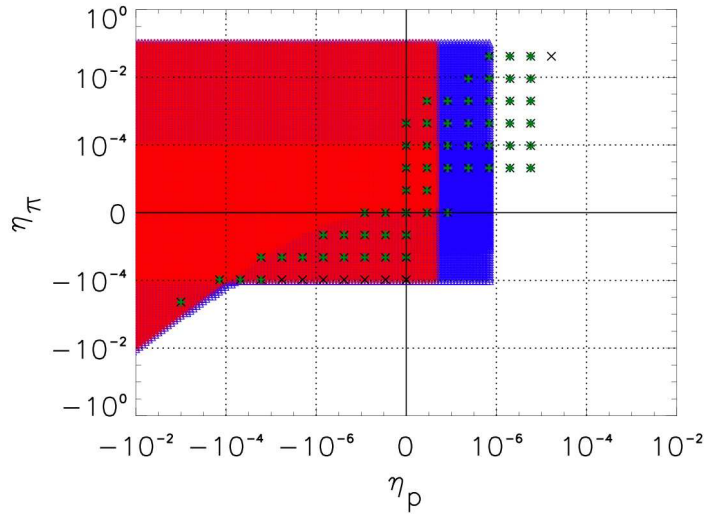}
   \caption{This plot shows the ($\eta_{p}$,$\eta_{\pi}$) parameter space allowed by different UHECR observations. The red and blue shaded regions correspond to the portion of parameter space for which the energy threshold for VC emission is higher than, respectively, $10^{20.25}~\eV$ and $10^{19.95}~\eV$, so that it does not conflict with PAO observations. The green circles and black crosses represent points in the parameter space for which LV effects in the UHECR spectrum are still in agreement with experimental data. They correspond respectively to an agreement with data within $2\sigma$ and $3\sigma$ CL.}
   \label{fig:VC}
\end{figure}

We consider first of all the fact that protons with energy $\gtrsim 10^{20}~\eV$ have been observed. A straightforward constraint is then implied by the fact that these protons do not lose a significant amount of their energy through VC emission during propagation. In order to be able to reproduce the highest energy point of AUGER data (whose energy is about $10^{20.25}~\eV$), we are forced to demand
\begin{equation}
E_{\rm th}^{\rm VC} > 10^{20.25}~\eV\;.
\label{eq:constraintVC}
\end{equation}
Photon VC emission does not depend on the pion LV coefficient, but it may happen only if $\eta_{p}> 0$, according to Eq.~(\ref{eq:pgVC}), hence it places a limit only on $\eta_{p}>0$. On the other hand, for some combinations of $(\eta_{p},\eta_{\pi})$ pion VC emission may become the dominant energy loss channel for UHECR protons. The portion of parameter space allowed by Eq.~\ref{eq:constraintVC} is the red region in Fig.~\ref{fig:VC}. However, this constraint is not as robust as it would seem at first sight, as the measured flux at this energy is compatible with 0 at $2\sigma$ Confidence Level (CL). From this point of view, it is safer to place a VC constraint at a slightly lower energy than the maximum one. We decide then to consider as our reference energy $10^{19.95}~\eV$, which corresponds to the highest energy AUGER observation which is not compatible with 0 at $3\sigma$. The constraint obtained in this way is shown as the blue region in Fig.~\ref{fig:VC}.

Further tightening of this region might be achieved by considering
modifications of the GZK reaction. We will neglect in the following the region $(\eta_{p}>0,\eta_{\pi}<0)$, which is strongly constrained by VC, and we run MonteCarlo simulations in the region
\begin{eqnarray}
\nonumber 10^{-8} < &|\eta_{p}|& < 10^{-3} \\
\nonumber 10^{-6} < &|\eta_{\pi}|& < 4\;.
\end{eqnarray}
We also consider the lines $\eta_{p} = 0$ and $\eta_{\pi} = 0$.

A $\chi^{2}$ strategy would seem most suitable in order to check different ($\eta_{p}$,$\eta_{\pi}$) LV models against experimental data. Data are taken from \cite{Roth:2007in}.
It is interesting to notice that there are values of the pair
$(\eta_p,\eta_\pi)$ that provide a better fit of data than the LI
model. In particular, the minimum of the $\chi^{2}$ ($\chi^{2}_{\rm
min}= 1.45$) occurs for $(\eta_p,\eta_\pi)\sim (2.4\times10^{-7},
9.5\times10^{-5})$, while the $\chi^{2}$ associated to the LI
propagation model is of the order of 6.8.  However, since we have
more parameters available one would expect such a lowering of the
$\chi^2$ value.  Only major progress in both theoretical and experimental understanding of the UHECR spectrum could lead to better discrimination between LI and LV best-fit models. 

Using the best fit value of the $\chi^{2}$, constraints at 95\% and
99\% CL can be placed, respectively, at $\chi^2 > 7.4$ and $\chi^2
> 10.6$ (see \cite{pdg} for further details). The green circles and
black crosses in Fig.~\ref{fig:VC} represent points in the parameter
space allowed at 95\% and 99\% CL respectively. We notice that there
is no allowed point in the quadrant $(\eta_{p}<0, \eta_{\pi}>0)$. In
fact, the recovery feature we found in this region of the parameter
space is so strong that even the smallest values of the LV
parameters we considered $(\eta_{p} = -10^{-8}, \eta_{\pi} =
10^{-6})$ produce UHECR spectra incompatible with data.

Summarizing, the final constraints implied by UHECR physics are (at
99\% CL)
\begin{eqnarray}
\nonumber
-10^{-3} \lesssim &\eta_{p}& \lesssim 10^{-6}\\
-10^{-3} \lesssim &\eta_{\pi}&  \lesssim 10^{-1} \qquad (\eta_{p} > 0)\\
\nonumber & & \lesssim 10^{-6} \qquad (\eta_{p} < 0)\;.
\label{eq:finalconstraint}
\end{eqnarray}
As it can be noticed, the constraint $\eta_{p} \gtrsim -10^{-3}$ is placed at one of the edges of our simulation field. This is due to the fact that for $\eta_{p}\simeq \eta_{\pi}\simeq -10^{-3}$ protons of energy above $10^{19.85}~\eV$ lose energy dramatically in pion production, while below this energy they do not effectively interact with the radiation backgrounds. The combination of these two effects yields a GZK-like feature, in statistical agreement with data\footnote{It is interesting to note that the situation in which both coefficients are negative and equal is envisaged in other frameworks of LV, such as \cite{Ellis:2008gg}. However, due to renormalization group flow this equality, even if realized at the QG scale, would not generically hold for UHECR energy scales without an {\em ad hoc} symmetry or other mechanism to protect it.}. We checked, however, that for more negative values of the LV parameters this effect happens at too low energy to be compatible with data. Hence, this constraint is robust.

\section{Conclusions}

In this work we have investigated the consequences of relaxing the
assumption of Lorentz invariance in the physics of UHECRs. Motivated
by naturalness arguments we focused on a particular realisation of
LV in which it is described by the addition, in an EFT context, of
mass dimension five and six CPT even operators to the Standard Model
Lagrangian.

A careful analysis of the physics intervening in the propagation of
protons with energy $E_{p} > {few}\times 10^{19}~\eV$ allowed us to
identify how LV would modify the arriving spectrum of UHECRs at
Earth. Due to photo-hadronic interactions with CMB photons, the
spectrum of UHECRs is expected to be suppressed above a certain
energy, corresponding to the threshold energy at which the GZK
process becomes effective. The strength of the suppression depends
upon physical uncertainties about the UHECR sources, such as the
distance of the closest source from Earth (because the
mean-free-path of protons for such a process is of the order of few
Mpc). However, we found that the effect of LV is not degenerate with
this uncertainty, and can give rise to a very distinct signature
entirely unexpected in the LI case. A detailed observation of the
GZK cut-off may  therefore, in principle, be used to probe the
presence of LV effects at these energies, e.g.~through the observation of a recovery of the spectrum at high energies.

Moreover, we are able to generalize and to strengthen the
constraints on $\eta_{p}$ and $\eta_{\pi}$ compared to previous
works. On the one hand we considered the full parameter space, with
only one simplifying assumption, parity, on the LV coefficients. On
the other hand, we placed robust constraints, through a careful
statistical analysis of the agreement between model expectations and
observational data, strengthening by more than four orders of
magnitude previous limits in some regions of the parameter space.

However, this analysis also shows that significant improvements on
constraints of LV obtained using this method will be possible only
when better data becomes available. Improvements on both statistics
and energy resolution of data at energies $E > 10^{19.6}~\eV$ are
definitely needed to achieve this.

\section*{Acknowledgements}
LM acknowledges support from SISSA during the early stages of
preparation of this work.

\begin{appendix}

\section{Simple Analytic Form for UHECR Proton Attenuation}
\label{analytic}

\begin{figure}[tbp]
\centering\leavevmode
\includegraphics[width=0.4\linewidth,angle=-90]{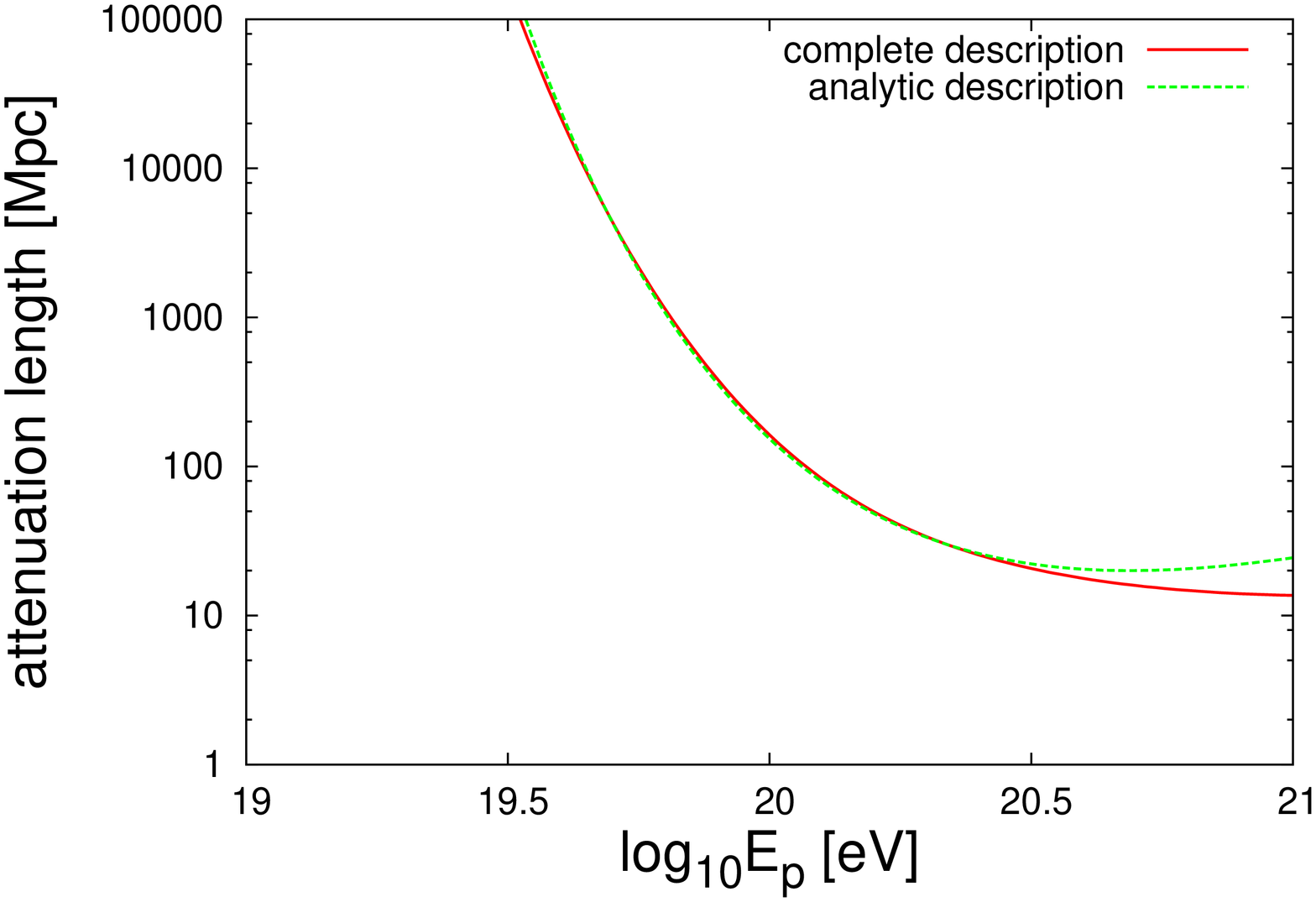}
\caption{The proton attenuation length as a function of proton
energy demonstrating the effectiveness of the analytic description
for protons below 10$^{20.5}$~eV.} \label{fig:horizon}
\end{figure}

Assuming that the $p\gamma$ interaction occurs predominantly at the
onset of the $\Delta$-resonance ($p\gamma\rightarrow\Delta^{+}\rightarrow p\pi^{0}/n\pi^{+}$), whose width is assumed to be $\Delta_{p,\gamma}$, we can write the
interaction rate given in Eq.~(\ref{mean-free-path}) as
\begin{eqnarray}
ct_{p,\gamma}^{-1}=\sigma_{p,\gamma}\int_{\frac{E_{p,\gamma}-\Delta_{p,\gamma}}{2\Gamma}}^{\frac{E_{p,\gamma}+\Delta_{p,\gamma}}{2\Gamma}}n(E)dE\;,
\label{eq:horizon}
\end{eqnarray}
where $E_{p,\gamma}=310~\MeV$ is the photon threshold energy in the
proton rest frame corresponding to the $\Delta-$resonance,
$\Delta_{p,\gamma}=100~\MeV$ is the width of the $\Delta$-resonance, $n(E)$ is the photon number energy distribution and $\sigma_{p,\gamma}\simeq 0.5$~mb is the interaction cross section.
Assuming that $n(E)$ corresponds to the CMB spectrum, at a
temperature $T=2.73$~K, Eq.~(\ref{eq:horizon}) can be re-written as,
\begin{eqnarray}
ct_{p,\gamma}^{-1}=\sigma_{p,\gamma}
n_{\gamma}\int_{x_{0}}^{x_{1}}f(x)dx
\end{eqnarray}
where $f(x)=x^{2}/(e^{x}-1)$, $x_{0}=E_{p,0}/(3E_{p})$,
$x_{1}=2E_{p,0}/(3E_{p})=2x_{0}$, and
$E_{p,0}=m_{p}E_{p,\gamma}/kT=10^{20.6}$~eV. Since at threshold
$E_{p,th}\sim m_{p}m_{\pi}/2E_{\gamma}=10^{20}$~eV, at threshold the
integral probes the $x\approx10$ region. With the inelasticity of
these collisions being roughly 20\%, the corresponding attenuation
lengths are
\begin{eqnarray}
l_{\rm horiz.}=\frac{l_{0}}{\left[e^{-x}(1-e^{-x})\right]}
\label{eq:horizon-rough}
\end{eqnarray}
where $l_{0}=5$~Mpc, $x=E_{p,0}/3E_{p}$ and
$E_{p,0}/3=10^{20.53}$~eV. Equation (\ref{eq:horizon-rough}) is
represented in Fig.~\ref{fig:horizon}, where it is compared to the
results of a full numerical computation of the GZK horizon.

\end{appendix}

\section*{References}
%%%%%%%%%%%%%%%%%%%%%%%%%%%%%%%%%%
%\bibliographystyle{apsrev}

%%%%%%%%%%%%%%%%%%

\begin{thebibliography}{999}

\bibitem{KS89}
V.~A.~Kostelecky and S.~Samuel,
%``Spontaneous Breaking Of Lorentz Symmetry In String Theory,''
Phys.\ Rev.\ D {\bf 39}, 683 (1989).
%%CITATION = PHRVA,D39,683;%%

%\cite{Ellis:2008gg}
\bibitem{Ellis:2008gg}
  J.~R.~Ellis, N.~E.~Mavromatos and D.~V.~Nanopoulos,
  %``Derivation of a Vacuum Refractive Index in a Stringy Space-Time Foam
  %Model,''
  Phys.\ Lett.\  B {\bf 665}, 412 (2008)
  [arXiv:0804.3566 [hep-th]].
  %%CITATION = PHLTA,B665,412;%%

\bibitem{Horava:2009uw}
  P.~Horava,
  %``Quantum Gravity at a Lifshitz Point,''
  arXiv:0901.3775 [hep-th].
  %%CITATION = ARXIV:0901.3775;%%

%------------------------------
%\cite{Gambini:1998it}
\bibitem{LoopQG}
R.~Gambini and J.~Pullin,
%``Nonstandard optics from quantum spacetime,''
Phys.\ Rev.\ D {\bf 59}, 124021 (1999).
%%[arXiv:gr-qc/9809038]

%\cite{Rovelli:2002vp}
\bibitem{Rovelli:2002vp}
  C.~Rovelli and S.~Speziale,
  %``Reconcile Planck-scale discreteness and the Lorentz-Fitzgerald
  %contraction,''
  Phys.\ Rev.\  D {\bf 67}, 064019 (2003)
  [arXiv:gr-qc/0205108].
  %%CITATION = PHRVA,D67,064019;%%

\bibitem{Alfaro:2002ya}
  J.~Alfaro and G.~Palma,
  %``Loop quantum gravity and ultra high energy cosmic rays,''
  Phys.\ Rev.\  D {\bf 67} (2003) 083003
  [arXiv:hep-th/0208193].
  %%CITATION = PHRVA,D67,083003;%%


%\cite{Carroll:2001ws}
\bibitem{Carroll:2001ws}
S.~M.~Carroll, J.~A.~Harvey, V.~A.~Kostelecky, C.~D.~Lane and
T.~Okamoto,
%``Noncommutative field theory and Lorentz violation,''
Phys.\ Rev.\ Lett.\ {\bf 87}, 141601 (2001).
%[arXiv:hep-th/0105082].
%%CITATION = HEP-TH 0105082;%%

\bibitem{Lukierski:1993wx}
  J.~Lukierski, H.~Ruegg and W.~J.~Zakrzewski,
  %``Classical Quantum Mechanics Of Free Kappa Relativistic Systems,''
  Annals Phys.\  {\bf 243} (1995) 90
  [arXiv:hep-th/9312153].
  %%CITATION = APNYA,243,90;%%

\bibitem{AmelinoCamelia:1999pm}
  G.~Amelino-Camelia and S.~Majid,
  %``Waves on noncommutative spacetime and gamma-ray bursts,''
  Int.\ J.\ Mod.\ Phys.\  A {\bf 15} (2000) 4301
  [arXiv:hep-th/9907110].
  %%CITATION = IMPAE,A15,4301;%%

%\cite{Chaichian:2004za}
\bibitem{Chaichian:2004za}
  M.~Chaichian, P.~P.~Kulish, K.~Nishijima and A.~Tureanu,
  %``On a Lorentz-invariant interpretation of noncommutative space-time and  its
  %implications on noncommutative QFT,''
  Phys.\ Lett.\  B {\bf 604}, 98 (2004)
  [arXiv:hep-th/0408069].
  %%CITATION = PHLTA,B604,98;%%

%\cite{Amelino-Camelia:1997gz}
\bibitem{AmelinoCamelia:1997gz}
G.~Amelino-Camelia, J.~R.~Ellis, N.~E.~Mavromatos, D.~V.~Nanopoulos
and S.~Sarkar,
%``Potential Sensitivity of Gamma-Ray Burster Observations to Wave Dispersion in Vacuo,''
Nature {\bf 393}, 763 (1998).
%[arXiv:astro-ph/9712103].
%%CITATION = ASTRO-PH 9712103;%%

%\cite{Burgess:2002tb}
\bibitem{Burgess:2002tb}
C.~P.~Burgess, J.~Cline, E.~Filotas, J.~Matias and G.~D.~Moore,
%``Loop-generated bounds on changes to the graviton dispersion relation,''
JHEP {\bf 0203}, 043 (2002).
% [arXiv:hep-ph/0201082].
%%CITATION = HEP-PH 0201082;%%

%-----------------------------------------------------------------
\bibitem{Analogues}
C.~Barcelo, S.~Liberati and M.~Visser,
 {\em ``Analogue gravity,''}
  Living Rev.\ Rel.\  {\bf 8}, 12 (2005).
 [arXiv:gr-qc/0505065].
  %%CITATION = GR-QC 0505065;%%

%\cite{Burgess:2003jk}
\bibitem{Burgess:2003jk}
  C.~P.~Burgess,
  %``Quantum gravity in everyday life: General relativity as an effective  field
  %theory,''
  Living Rev.\ Rel.\  {\bf 7}, 5 (2004)
  [arXiv:gr-qc/0311082].
  %%CITATION = 00222,7,5;%%

\bibitem{add}
%\cite{Arkani-Hamed:1998rs}
%%%\bibitem{Arkani-Hamed:1998rs}
N. Arkani-Hamed, S. Dimopoulos and G. Dvali,
%{\it The hierarchy problem and new dimensions at a millimeter},
Phys.\ Lett.\ B {\bf 429} (1998) 263;
%[arXiv:hep-ph/9803315].
%%CITATION = HEP-PH 9803315.%%
%``Phenomenology, astrophysics and cosmology of theories with
%sub-millimeter dimensions and TeV scale quantum gravity,'' Phys.\
%Rev.\ D {\bf 59} (1999) 086004;
% [arXiv:hep-ph/9807344].
%%CITATION = HEP-PH 9807344;%%
L. Randall and R. Sundrum,
%{\it A large mass hierarchy from a small extra dimension},
Phys.\ Rev.\ Lett.\  {\bf 83} (1999) 3370;
%[arXiv:hep-ph/9905221].
%%CITATION = HEP-PH 9905221.%%
%``An alternative to compactification,''
Phys.\ Rev.\ Lett.\  {\bf 83} (1999) 4690.
%[arXiv:hep-th/9906064].
%%CITATION = HEP-TH 9906064;%%

%\cite{Han:2004wt}
\bibitem{Han:2004wt}
  T.~Han and S.~Willenbrock,
  %``Scale of quantum gravity,''
  Phys.\ Lett.\  B {\bf 616}, 215 (2005)
  [arXiv:hep-ph/0404182].
  %%CITATION = PHLTA,B616,215;%%

\bibitem{Mattingly:2005re}
D.~Mattingly,
%  {\em ``Modern tests of Lorentz invariance,''}
  Living Rev.\ Rel.\  {\bf 8}, 5 (2005)
%  [arXiv:gr-qc/0502097].
  %%CITATION = GR-QC 0502097;%%

\bibitem{AmelinoCamelia:2002dx}
  G.~Amelino-Camelia,
  %``Phenomenology of Planck-scale Lorentz-symmetry test theories,''
  New J.\ Phys.\  {\bf 6} (2004) 188
  [arXiv:gr-qc/0212002].
  %%CITATION = NJOPF,6,188;%%

\bibitem{Bertolami:2000qa}
  O.~Bertolami,
  %``Ultra-high energy cosmic rays and symmetries of spacetime,''
  Gen.\ Rel.\ Grav.\  {\bf 34} (2002) 707
  [arXiv:astro-ph/0012462].
  %%CITATION = GRGVA,34,707;%%
  
\bibitem{AmelinoCamelia:2008qg}
  G.~Amelino-Camelia,
  %``Quantum Gravity Phenomenology,''
  arXiv:0806.0339 [gr-qc].
  %%CITATION = ARXIV:0806.0339;%%
  
%\cite{Colladay:1998fq}
\bibitem{Colladay:1998fq}
  D.~Colladay and V.~A.~Kostelecky,
  %``Lorentz-violating extension of the standard model,''
  Phys.\ Rev.\  D {\bf 58}, 116002 (1998)
  [arXiv:hep-ph/9809521].
  %%CITATION = PHRVA,D58,116002;%%

\bibitem{Myers:2003fd}
  R.~C.~Myers and M.~Pospelov,
  %``Experimental challenges for quantum gravity,''
  Phys.\ Rev.\ Lett.\  {\bf 90} (2003) 211601
  [arXiv:hep-ph/0301124].
  %%CITATION = PRLTA,90,211601;%%

%\cite{Bolokhov:2007yc}
\bibitem{Bolokhov:2007yc}
  P.~A.~Bolokhov and M.~Pospelov,
  %``Classification of dimension 5 Lorentz violating interactions in the
  %standard model,''
  Phys.\ Rev.\  D {\bf 77}, 025022 (2008)
  [arXiv:hep-ph/0703291].
  %%CITATION = PHRVA,D77,025022;%%

%\cite{Mattingly:2008pw}
\bibitem{Mattingly:2008pw}
  D.~Mattingly,
  %``Have we tested Lorentz invariance enough?,''
  arXiv:0802.1561 [gr-qc].
  %%CITATION = ARXIV:0802.1561;%%

%\cite{GrootNibbelink:2004za}
\bibitem{GrootNibbelink:2004za}
  S.~Groot Nibbelink and M.~Pospelov,
  %``Lorentz violation in supersymmetric field theories,''
  Phys.\ Rev.\ Lett.\  {\bf 94}, 081601 (2005)
  [arXiv:hep-ph/0404271].
  %%CITATION = PRLTA,94,081601;%%

%\cite{Collins:2004bp}
\bibitem{Collins:2004bp}
  J.~Collins, A.~Perez, D.~Sudarsky, L.~Urrutia and H.~Vucetich,
  %``Lorentz invariance: An additional fine-tuning problem,''
  Phys.\ Rev.\ Lett.\  {\bf 93}, 191301 (2004)
  [arXiv:gr-qc/0403053].
  %%CITATION = PRLTA,93,191301;%%

\bibitem{Galaverni:2007tq}
  M.~Galaverni and G.~Sigl,
  %``Lorentz Violation in the Photon Sector and Ultra-High Energy Cosmic Rays,''
  Phys. Rev. Lett., {\bf 100}, 021102 (2008).
  arXiv:0708.1737 [astro-ph].
  %%CITATION = ARXIV:0708.1737;%%

\bibitem{Maccione:2008iw}
  L.~Maccione and S.~Liberati,
  %``GZK photon constraints on Planck scale Lorentz violation in QED,''
  arXiv:0805.2548 [astro-ph].
  %%CITATION = ARXIV:0805.2548;%%

\bibitem{Galaverni:2008yj}
  M.~Galaverni and G.~Sigl,
  %``Lorentz Violation and Ultrahigh-Energy Photons,''
  Phys.\ Rev.\  D {\bf 78} (2008) 063003
  [arXiv:0807.1210 [astro-ph]].
  %%CITATION = PHRVA,D78,063003;%%

\bibitem{Aloisio:2000cm}
  R.~Aloisio, P.~Blasi, P.~L.~Ghia and A.~F.~Grillo,
  %``Probing the structure of space-time with cosmic rays,''
  Phys.\ Rev.\  D {\bf 62} (2000) 053010
  [arXiv:astro-ph/0001258].
  %%CITATION = PHRVA,D62,053010;%%

\bibitem{Stecker:2004xm}
  F.~W.~Stecker and S.~T.~Scully,
  %``Lorentz invariance violation and the spectrum and source power of
  %ultrahigh energy cosmic rays,''
  Astropart.\ Phys.\  {\bf 23} (2005) 203
  [arXiv:astro-ph/0412495].
  %%CITATION = APHYE,23,203;%%

\bibitem{GonzalezMestres:2009di}
  L.~Gonzalez-Mestres,
  %``AUGER-HiRes results and models of Lorentz symmetry violation,''
  arXiv:0902.0994 [astro-ph.HE].
  %%CITATION = ARXIV:0902.0994;%%
  
%\cite{Scully:2008jp}
\bibitem{Scully:2008jp}
  S.~T.~Scully and F.~W.~Stecker,
  %``Lorentz Invariance Violation and the Observed Spectrum of Ultrahigh Energy
  %Cosmic Rays,''
  arXiv:0811.2230 [astro-ph].
  %%CITATION = ARXIV:0811.2230;%%

\bibitem{Bi:2008yx}
  X.~J.~Bi, Z.~Cao, Y.~Li and Q.~Yuan,
  %``Testing Lorentz Invariance with Ultra High Energy Cosmic Ray Spectrum,''
  arXiv:0812.0121 [astro-ph].
  %%CITATION = ARXIV:0812.0121;%%

\bibitem{Glinka:2008tr}
  L.~A.~Glinka,
  %``On some consequences of the Snyder-Sidharth deformation of Special
  %Relativity,''
  arXiv:0812.0551 [hep-th].
  %%CITATION = ARXIV:0812.0551;%%
  
\bibitem{Jacobson:2002hd}
  T.~Jacobson, S.~Liberati and D.~Mattingly,
  %``Threshold effects and Planck scale Lorentz violation: Combined constraints
  %from high energy astrophysics,''
  Phys.\ Rev.\  D {\bf 67} (2003) 124011
  [arXiv:hep-ph/0209264].
  %%CITATION = PHRVA,D67,124011;%%

%\cite{Collaboration:2008ih}
\bibitem{Collaboration:2008ih}
  T.~I.~Collaboration,
  %``Search for Point Sources of High Energy Neutrinos with Final Data from
  %AMANDA-II,''
  arXiv:0809.1646 [astro-ph].
  %%CITATION = ARXIV:0809.1646;%%

\bibitem{Gagnon:2004xh}
  O.~Gagnon and G.~D.~Moore,
  %``Limits on Lorentz violation from the highest energy cosmic rays,''
  Phys.\ Rev.\  D {\bf 70} (2004) 065002
  [arXiv:hep-ph/0404196].
  %%CITATION = PHRVA,D70,065002;%%

\bibitem{Gaisser:2006sf}
  T.~K.~Gaisser and T.~Stanev,
  %``High-Energy Cosmic Rays,''
  Nucl.\ Phys.\  A {\bf 777} (2006) 98.
  %%CITATION = NUPHA,A777,98;%%

  \bibitem{Bergman:2007kn}
  D.~R.~Bergman and J.~W.~Belz,
  %``Cosmic rays: The second knee and beyond,''
  J.\ Phys.\ G {\bf 34}, R359 (2007)
  [arXiv:0704.3721 [astro-ph]].
  %%CITATION = JPHGB,G34,R359;%%

\bibitem{Hooper:2006tn}
  D.~Hooper, S.~Sarkar and A.~M.~Taylor,
  %``The intergalactic propagation of ultra-high energy cosmic ray nuclei,''
  Astropart.\ Phys.\  {\bf 27} (2007) 199
  [arXiv:astro-ph/0608085].
  %%CITATION = APHYE,27,199;%%

\bibitem{Cronin:2007zz}
  J.~Abraham {\it et al.}  [Pierre Auger Collaboration],
  %``Correlation of the highest energy cosmic rays with nearby extragalactic
  %objects,''
  Science {\bf 318} (2007) 938
  [arXiv:0711.2256 [astro-ph]].
  %%CITATION = SCIEA,318,938;%%

\bibitem{Ahlers:2005sn}
  M.~Ahlers, L.~A.~Anchordoqui, H.~Goldberg, F.~Halzen, A.~Ringwald and T.~J.~Weiler,
  %``Neutrinos as a diagnostic of cosmic ray galactic / extra-galactic
  %transition,''
  Phys.\ Rev.\  D {\bf 72} (2005) 023001
  [arXiv:astro-ph/0503229].
  %%CITATION = PHRVA,D72,023001;%%

\bibitem{Ahlers:2009rf}
  M.~Ahlers, L.~A.~Anchordoqui and S.~Sarkar,
  %``Neutrino diagnostics of ultra-high energy cosmic ray protons,''
  arXiv:0902.3993 [astro-ph.HE].
  %%CITATION = ARXIV:0902.3993;%%
  
\bibitem{Allard:2005cx}
 D.~Allard, E.~Parizot and A.~V.~Olinto,
 %``On the transition from Galactic to extragalactic cosmic-rays: spectral and
 %composition features from two opposite scenarios,''
 Astropart.\ Phys.\  {\bf 27}, 61 (2007)
 [arXiv:astro-ph/0512345].
 %%CITATION = APHYE,27,61;%% 

\bibitem{Berezinsky:2007wf}
 V.~Berezinsky,
 %``Transition from galactic to extragalactic cosmic rays,''
 arXiv:0710.2750 [astro-ph].
 %%CITATION = ARXIV:0710.2750;%%

\bibitem{Unger:2007mc}
  M.~Unger  [The Pierre Auger Collaboration],
  %``Study of the Cosmic Ray Composition above 0.4 EeV using the Longitudinal
  %Profiles of Showers observed at the Pierre Auger Observatory,''
  arXiv:0706.1495 [astro-ph].
  %%CITATION = ARXIV:0706.1495;%%

\bibitem{gzk}
  K.~Greisen,
  %``End To The Cosmic Ray Spectrum?,''
  Phys.\ Rev.\ Lett.\  {\bf 16} (1966) 748;   Zatsepin, Kuzmin, Sov.Phys.JETP 4, 78.
  %%CITATION = PRLTA,16,748;%%

\bibitem{Abbasi:2007sv}
  R.~Abbasi {\it et al.}  [HiRes Collaboration],
  %``Observation of the GZK cutoff by the HiRes experiment,''
  arXiv:astro-ph/0703099.
  %%CITATION = ASTRO-PH/0703099;%%

\bibitem{Roth:2007in}
  M.~Roth  [Pierre Auger Collaboration],
  %``Measurement of the UHECR energy spectrum using data from the Surface
  %Detector of the Pierre Auger Observatory,''
  arXiv:0706.2096 [astro-ph].
  %%CITATION = ARXIV:0706.2096;%%

\bibitem{Hooper:2008pm}
  D.~Hooper, S.~Sarkar and A.~M.~Taylor,
  %``The Intergalactic Propagation of Ultra-High Energy Cosmic Ray Nuclei: An
  %Analytic Approach,''
  Phys.\ Rev.\  D {\bf 77} (2008) 103007
  [arXiv:0802.1538 [astro-ph]].
  %%CITATION = PHRVA,D77,103007;%%

\bibitem{Allard:2008gj}
 D.~Allard, N.~G.~Busca, G.~Decerprit, A.~V.~Olinto and E.~Parizot,
 %``Implications of the cosmic ray spectrum for the mass composition at the
 %highest energies,''
 JCAP {\bf 0810} (2008) 033
 [arXiv:0805.4779 [astro-ph]].
 %%CITATION = JCAPA,0810,033;%%

\bibitem{Mucke:1999yb}
  A.~Mucke, R.~Engel, J.~P.~Rachen, R.~J.~Protheroe and T.~Stanev,
  %``Monte Carlo simulations of photohadronic processes in astrophysics,''
  Comput.\ Phys.\ Commun.\  {\bf 124} (2000) 290
  [arXiv:astro-ph/9903478].
  %%CITATION = CPHCB,124,290;%%

\bibitem{Mattingly:2002ba}
  D.~Mattingly, T.~Jacobson and S.~Liberati,
  %``Threshold configurations in the presence of Lorentz violating dispersion
  %relations,''
  Phys.\ Rev.\  D {\bf 67} (2003) 124012
  [arXiv:hep-ph/0211466].
  %%CITATION = PHRVA,D67,124012;%%


\bibitem{Stecker:1968uc}
  F.~W.~Stecker,
  %``Effect of photomeson production by the universal radiation field on
  %high-energy cosmic rays,''
  Phys.\ Rev.\ Lett.\  {\bf 21} (1968) 1016.
  %%CITATION = PRLTA,21,1016;%%

\bibitem{Jacobson:2005bg}
  T.~Jacobson, S.~Liberati and D.~Mattingly,
  %``Lorentz violation at high energy: Concepts, phenomena and astrophysical
  %constraints,''
  Annals Phys.\  {\bf 321} (2006) 150
  [arXiv:astro-ph/0505267].
  %%CITATION = ASTRO-PH 0505267;%%

\bibitem{Berezinsky:2002nc}
  V.~Berezinsky, A.~Z.~Gazizov and S.~I.~Grigorieva,
  %``On astrophysical solution to ultra high energy cosmic rays,''
  Phys.\ Rev.\  D {\bf 74} (2006) 043005
  [arXiv:hep-ph/0204357].
  %%CITATION = PHRVA,D74,043005;%%


%--------------------------------------------------------------------

%\bibitem{Maccione:2007yc}
%  L.~Maccione, S.~Liberati, A.~Celotti and J.~G.~Kirk,
%  %``New constraints on Planck-scale Lorentz Violation in QED from the Crab
%  %Nebula,''
%  JCAP {\bf 0710} (2007) 013
%  [arXiv:0707.2673 [astro-ph]].
%  %%CITATION = JCAPA,0710,013;%%


%\bibitem{Crab-pol}
 % L.~Maccione, S.~Liberati, A.~Celotti, J.~G.~Kirk and P.~Ubertini,
  %``
   %''
%in preparation

%\bibitem{Fan:2007zb}
%  Y.~Z.~Fan, D.~M.~Wei and D.~Xu,
%  %``Gamma-ray Burst UV/optical afterglow polarimetry as a probe of Quantum
%  %Gravity,''
%  Mon.\ Not.\ Roy.\ Astron.\ Soc.\  {\bf 376} (2006) 1857
%  [arXiv:astro-ph/0702006].
%  %%CITATION = MNRAA,376,1857;%%

%  \bibitem{Globus:2007bi}
%  N.~Globus, D.~Allard and E.~Parizot,
%  %``Propagation of high-energy cosmic rays in extragalactic turbulent magnetic
%  %fields: resulting energy spectrum and composition,''
%  arXiv:0709.1541 [astro-ph].
%  %%CITATION = ARXIV:0709.1541;%%

%  \bibitem{Aloisio:2006wv}
%  R.~Aloisio, V.~Berezinsky, P.~Blasi, A.~Gazizov, S.~Grigorieva and B.~Hnatyk,
%  %``A dip in the UHECR spectrum and the transition from galactic to
%  %extragalactic cosmic rays,''
%  Astropart.\ Phys.\  {\bf 27} (2007) 76
%  [arXiv:astro-ph/0608219].
%  %%CITATION = APHYE,27,76;%%

%
\bibitem{pdg}
W.-M. Yao\ W~M {\it et al.}~[The Particle Data Group]\ 2006  {\it
J. Phys.}\  {\bf G 33}   1

%%%%%%%%%%%%%%%%%%%%
\end{thebibliography}
\end{document}